\documentclass[a4paper]{article}
\usepackage{Odyssey2018}
\usepackage{epsfig,amssymb,amsmath}
\usepackage{bm}
\usepackage{empheq}
\usepackage{hyperref}

\hypersetup{
   bookmarks=true,         
   unicode=false,          
   pdftoolbar=true,        
   pdfmenubar=true,        
   pdffitwindow=false,     
   pdfstartview={FitH},    
   pdftitle={My title},    
   pdfauthor={Author},     
   pdfsubject={Subject},   
   pdfcreator={Creator},   
   pdfproducer={Producer}, 
   pdfkeywords={keyword1} {key2} {key3}, 
   pdfnewwindow=true,      
   colorlinks=true,       
   linkcolor=blue,          
   citecolor=blue,        
   filecolor=magenta,      
   urlcolor=blue           
}

\newcommand{\ts}{\textsuperscript}

\graphicspath{{figures/}}

\newcommand{\ie}{\textit{i}.\textit{e}., }
\newcommand{\eg}{\textit{e}.\textit{g}. }
\renewcommand{\vec}[1]{\bm{#1}}
\newcommand{\R}{\mathbb{R}}
\newcommand{\T}{^\intercal}
\DeclareMathOperator{\Tr}{tr\!}

\newcommand*\textfbox[2][Title]{%
  \begin{tabular}[b]{@{}c@{}}\footnotesize\textbf{#1}\\\fbox{#2}\end{tabular}}

\ninept

\title{Supervector Compression Strategies to Speed up I-Vector System Development}

\name{Ville Vestman, Tomi Kinnunen\thanks{This work was partially funded by Academy of Finland (project 309629).}
}

\address{
University of Eastern Finland, Finland \\
{\small \tt vvestman@cs.uef.fi, tkinnu@cs.uef.fi} }

\begin{document}

\maketitle

\begin{abstract}
The front-end factor analysis (FEFA), an extension of principal component analysis (PPCA) tailored to be used with Gaussian mixture models (GMMs), is currently the prevalent approach to extract compact utterance-level features (\mbox{i-vectors}) for automatic speaker verification (ASV) systems. Little research has been conducted comparing FEFA to the conventional PPCA applied to maximum a posteriori (MAP) adapted GMM supervectors.
We study several alternative methods, including PPCA, factor analysis (FA), and two supervised approaches, supervised PPCA (SPPCA) and the recently proposed probabilistic partial least squares (PPLS), to compress MAP-adapted GMM supervectors. The resulting i-vectors are used in ASV tasks with a probabilistic linear discriminant analysis (PLDA) back-end. We experiment on two different datasets, on the telephone condition of NIST SRE 2010 and on the recent VoxCeleb corpus collected from YouTube videos containing celebrity interviews recorded in various acoustical and technical conditions.
The results suggest that, in terms of ASV accuracy, the supervector compression approaches are on a par with FEFA. The supervised approaches did not result in improved performance. In comparison to FEFA, we obtained more than hundred-fold (100x) speedups in the total variability model (TVM) training using the PPCA and FA supervector compression approaches.

\end{abstract}

\section{Introduction}

Modern text-independent automatic speaker verification (ASV) relies heavily on the use of \emph{identity vectors} (i-vectors)~\cite{dehak2011front, kenny2012small}. \mbox{I-vectors} are compact representations of speech utterances containing useful information for speech-related classification tasks. 
The i-vector extraction pipeline involves many steps starting from the extraction of acoustic features such as \emph{Mel-frequency cepstral coefficients} (MFCCs), followed by the extraction of \emph{sufficient statistics} with the aid of an \emph{universal background model} (UBM), typically a \emph{Gaussian mixture model} (GMM)~\cite{reynolds2000speaker} or a \emph{deep neural network} (DNN) model~\cite{lei2014novel}. Sufficient statistics are then used to extract an i-vector, a fixed-length representation of an utterance, using a pre-trained \emph{total variability model} (TVM) that models the distribution of utterance-specific GMM supervectors.

The development and optimization of an i-vector based ASV system consists of a multiple time-consuming steps. The most notable ones are extraction of acoustic features and sufficient statistics, and training of UBM and TVM. The two former require processing of a large number of speech files and 
TVM training is among the most time consuming parts of the system development process. Thus, by reducing TVM training time, a meaningful positive effect to the total development time of ASV system can be achieved~\cite{glembek2011simplification,cumani2013memory}. This is particularly beneficial in studies focused on the acoustic front-end optimizations when one has to retrain the entire system when feature extractor is changed~\cite{vestman2017time}.

Previous studies on rapid i-vector extraction have primarily optimized computations in the standard \emph{front-end factor analysis} (FEFA) approach~\cite{dehak2011front, kenny2012small} by adopting new computational algorithms, often approximative in nature~\cite{cumani2013memory, cumani2014factorized, xu2018generalizing}. In this study, however, we focus on an alternative and straightforward compression of classic \emph{maximum a posteriori} (MAP)~\cite{reynolds2000speaker} adapted GMM supervectors with a goal of obtaining fast execution times without compromising on ASV accuracy.  
In fact, before FEFA, and its predecessor, \emph{joint factor analysis} (JFA)~\cite{kenny2007joint}, became prevalent, MAP adapted supervectors were commonly used with \emph{support vector machine} (SVM) to do speaker classification~\cite{campbell2006support}. Recently, however, the use of MAP adapted supervectors has been less common. 

Supervector compression, for example by using \emph{probabilistic principal component analysis} (PPCA)~\cite{roweis1998algorithms, tipping1999probabilistic}, provides a large computational saving in TVM estimation over FEFA~\cite{madikeri2014fast}. In FEFA, the posterior covariance matrix needed for i-vector extraction is \emph{utterance-dependent}, while in the supervector compression methods addressed in this study, covariance is shared among all speech utterances, which greatly reduces computation. As the TVM training set may consist of tens of thousands of utterances, the resulting computational saving is considerable~\cite{madikeri2014fast}.

The closest prior work similar in spirit to ours are \cite{madikeri2014fast} and~\cite{madikeri2012hybrid}, which we extend in many ways. In these two studies, the training of TVM is performed using PPCA. 
The acoustic feature vectors are \emph{hardly aligned} to a single UBM component. If they were \emph{softly aligned}, this approach would equal to using MAP adapted supervectors with a \emph{relevance factor} of 0~\cite{haris2012exploring}. Differing from~\cite{madikeri2014fast} and~\cite{madikeri2012hybrid},
we use MAP adapted supervectors to train TVM and we study how the choice of relevance factor affects the system performance.

Recently \cite{chen2017speaker}, TVM estimation using \emph{probabilistic partial least squares} (PPLS) was proposed as an alternative to FEFA. PPLS compresses supervectors in a \emph{supervised} way by taking advantage of the speaker labels in the training set. 
In the current study, we attempt to validate the positive results~\cite{chen2017speaker} obtained for Chinese mandarin corpus by using datasets containing English speech instead.
In~\cite{lei2010speaker}, supervision is added to the total variability matrix training by deploying \emph{supervised} PPCA (SPPCA)~\cite{yu2006supervised}. The SPPCA model is fundamentally the same as the PPLS model, with a difference in what has been used as the supervision data; PPLS has been used directly with speaker labels~\cite{chen2017speaker}, while SPPCA has been utilized with speaker-specific (not just utterance-specific) sufficient statistics~\cite{lei2010speaker}. In the current work, we study the use of SPPCA for supervector compression.

In addition to the above models, we adopt standard \emph{factor analysis} (FA) for supervector compression. Note, that this differs from the FEFA framework: In FEFA, the TVM training is based on the maximization of posterior probabilities of acoustic feature vectors, assumed to have been generated by a GMM~\cite{kenny2005eigenvoice}, whereas in FA (and PPCA), maximization is performed with respect to posterior probabilities of supervectors~\cite{bishop2006machine}.

We conduct comparisons of different methods in ASV setting using a recently released \emph{VoxCeleb} corpus~\cite{nagrani2017voxceleb}. The corpus contains ``real-world'' utterances obtained from YouTube videos of celebrity interviews using a fully automated data collection pipeline. The data is challenging for ASV due to large intra-speaker variability caused by large differences in environments, speaking styles, and technical conditions. In addition to VoxCeleb, we validate our findings with the telephone condition of NIST 2010 Speaker Recognition Evaluation corpus.

Our contributions can be summarized as follows. First, we present all the selected methods in an unified notation highlighting the important formulas regarding their implementation. Second, we aim at validating the results claimed in~\cite{chen2017speaker} regarding the recently proposed PPLS method on different corpora, and we extend the study by introducing the weighting scheme proposed in~\cite{lei2010speaker}. Third, we implement and test SPPCA in the supervector compression domain. Fourth, we compare all the methods in terms of ASV performances and training times of total variability models. Lastly, we propose a slight simplification to the maximization principle of TVM training.

\section{I-Vector Extraction by Front-End Factor Analysis}
\emph{Front-end factor analysis} (FEFA)~\cite{dehak2011front} is the current standard method for extracting utterance level features known as \emph{i-vectors}. In FEFA, a supervector $\vec{m}(u)$ of an utterance $u$ is modeled as
\[
	\vec{\widehat{m}}(u) = \vec{\mu} + V\vec{w}(u),
\]
where $\vec{\mu} \in \R^{h \times 1}$ is an utterance-independent bias supervector, $\vec{w}(u) \in \R^{d \times 1}$  is a low dimensional representation of an utterance supervector, \ie an i-vector, and $V \in \R^{h \times d}$ is a mapping between low and high dimensional spaces known as \emph{total variability matrix}. The mathematical foundation of FEFA is presented in detail in~\cite{kenny2005eigenvoice}.

The traditional way of TVM estimation and i-vector extraction begins with computing frame posterior probabilities for short-term spectral features (\eg MFCCs) of an utterance with each Gaussian component of UBM. These posteriors are then used to compute zeroth and first order sufficient statistics, 
\begin{align*}
n_c = \sum_{t=1}^T p_t(c), \\
\vec{f}_c = \sum_{t=1}^T p_t(c) \vec{x_t},
\end{align*}
where $\vec{x_t}$ is the $t$\ts{th} feature vector of the utterance and $p_t(c)$ is the posterior probability of $t$\ts{th} vector belonging to $c$\ts{th} component of UBM, computed with the aid of Bayes' rule.

Assuming that the prior distribution $p(\vec{w}(u))$ is stardard normal, it can be shown~\cite{kenny2005eigenvoice} that
\[
	p(\vec{w}(u) | X(u), V) = \mathcal{N}(\vec{\mu}(u), \Sigma(u)),
\]
where $X(u) = \{\vec{x}_1, \ldots, \vec{x}_T\}$ is a sequence of all feature vectors in the utterance $u$ and where
\begin{empheq}[box={\textfbox[I-Vector Extraction]}]{align*}
\Sigma(u) &= \left(I + \sum_{c=1}^C n_c(u) V_c\T \Sigma_c^{-1} V_c \right)^{-1}, \\
\vec{\mu}(u) &= \Sigma(u) \sum_{c=1}^C V_c\T \Sigma_c^{-1} (\vec{f}_c(u) - n_c(u)\vec{\mu}_c).
\end{empheq}
In the above equations, $\Sigma_c$ is the covariance matrix of the $c$\ts{th} UBM component, and $V_c$ and $\vec{\mu}_c$ are component-wise representations of $V$ and $\vec{\mu}$ so that
\[
V = \begin{bmatrix} V_1 \\ \vdots \\ V_C \end{bmatrix} \quad \textrm{ and } \quad \vec{\mu} = \begin{bmatrix} \vec{\mu}_1 \\ \vdots \\ \vec{\mu}_C \end{bmatrix},
\]
where the vectors $\vec{\mu}_1, \ldots, \vec{\mu}_C$ are the means of the UBM components. Mean $\vec{\mu}(u)$ of the posterior i-vector distribution is the i-vector of the utterance $u$.

The matrix $V$ is estimated using an offline training set of $U$ utterances by maximizing  
\begin{equation}
\label{kennyMaximization}
\sum_{u=1}^U \mathbb{E}[\ln p(X(u)| \vec{w}(u), V)],
\end{equation}
where the expectations are taken with respect to posterior i-vector distributions~\cite{kenny2005eigenvoice}. This leads to an update formula
\begin{empheq}[box={\textfbox[Update Formula for $V$]}]{align*}
&V_c \!=\! \left(\sum_{u=1}^U \vec{f}_c(u) \vec{\mu}(u)\T \!\right) \! \! \left( \sum_{u=1}^U n_c(u) \mathbb{E}_{\mu \mu}(u) \! \right)^{\!\!\!-1}, \\
&\mathbb{E}_{\mu \mu}(u) = \Sigma(u) + \vec{\mu}(u) \vec{\mu}(u)\T.
\end{empheq}
Training of $V$ is iterative; one iteration consists of computing $\Sigma(u)$, $\vec{\mu}(u)$, and $\mathbb{E}_{\mu \mu}(u)$ for all training utterances by keeping $V$ fixed and then updating $V$ using the computed values. During the first iteration, parameters of posterior distributions are computed using a randomly initialized matrix $V$.

\section{I-Vector Extraction by Supervector Compression}

\label{compressionSection}

In this section, we present multiple approaches to compressing MAP adapted GMM supervectors to low-dimensional representations, which we will also refer as ``i-vectors''. Unlike FEFA, these approaches do not assume the underlying speaker model to be GMM.

In relevance MAP, the adapted mean vectors $\vec{\hat \mu}_c$, \mbox{$c = 1,\ldots, C$}, for utterance's GMM are obtained from UBM by computing
\[
\vec{\hat \mu}_c = \alpha_c \vec{f}_c + (1-\alpha_c)\vec{\mu}_c,
\]
where 
\[
	\alpha_c = \frac{n_c}{n_c + r}
\]
and $r \geq 0$ is the \emph{relevance factor} to be optimized~\cite{reynolds2000speaker}. When $r \to 0$, then $\alpha_c \to 1$, and when $r = 0$, the mean vectors are solely determined by the sufficient statistics $\vec{f}_c$. If $r$ is large, then the adapted mean vectors remain close to UBM's mean vectors. Adapted mean vectors $\vec{\hat \mu}_c$ are concatenated together to form a \emph{supervector} for the utterance.

\subsection{Principal Component Analysis}
Being one of the most commonly used dimension reduction techniques, we include the conventional \emph{principal component analysis} (PCA)~\cite{pearson1901pca} as a baseline method for supervector compression. The PCA transformation matrix, consisting of eigenvectors of data covariance matrix, can be used to transform high dimensional supervectors to low dimensional i-vectors. In this study, we use the standard \emph{singular value decomposition} (SVD) approach for PCA computation. However, for high dimensional data, PCA could be computed more efficiently by adopting iterative PCA algorithms~\cite{roweis1998algorithms}.

\subsection{Probabilistic Principal Component Analysis}

\label{ppcaSection}

\emph{Probabilistic principal component analysis} (PPCA)~\cite{roweis1998algorithms, tipping1999probabilistic} models observations using a linear-Gaussian framework. In this section, we present a compact self-contained formulation of PPCA in the context of supervectors. The rest of the methods discussed in Section \ref{compressionSection} are formulated similarly and their theory can be easily formulated using PPCA as a starting point.

In PPCA, supervectors are modeled as
\begin{equation}
\label{ppcaModel}
	\vec{m}(u) = V\vec{w}(u) + \vec{\epsilon},
\end{equation}
where $\vec{w} \sim \mathcal{N}(\vec{0},I)$ and $\vec{\epsilon} \sim \mathcal{N}(\vec{0},\sigma^2 I)$. For brevity, we have omitted the bias term $\vec{\mu}$ from the right-hand side of the equation by assuming that supervectors have been centered using the data mean computed from the training data.

By using properties of normally distributed random variables and by assuming that $V$ is given, from \eqref{ppcaModel} it follows that
\[
	p(\vec{m}(u)|\vec{w}(u)) = \mathcal{N}(V \vec{w}(u), \sigma^2 I).
\]
Further, in Appendix \ref{posteriorProof}, we show that
\begin{equation}
\label{posteriorDistribution}
p(\vec{w}(u)|\vec{m}(u)) = \mathcal{N}(\vec{\mu}(u), \Sigma),
\end{equation}
where
\begin{empheq}[box={\textfbox[I-Vector Extraction]}]{align}
\label{ppcaExtraction}
\begin{split}
\Sigma &= \Big(I + \frac{1}{\sigma^2}V\T V\Big)^{-1}, \\
    \vec{\mu}(u) &= \frac{1}{\sigma^2} \Sigma V\T \vec{m}(u).
\end{split}
\end{empheq}
Unlike with FEFA, covariance $\Sigma$ of the posterior i-vector distribution does not depend on the utterance. Hence, by adopting PPCA instead of FEFA, the time complexity of computing the parameters of posterior distributions drops from \mbox{$O(U(CFd + Cd^2 + d^3))$} to $O(UCFd)$, where $F$ is the dimension of acoustic feature vectors~\cite{madikeri2014fast}.

In the current work, we study two different ways of obtaining $V$ for all the presented methods. The traditional approach (\textbf{max. principle 1}) maximizes
\begin{equation} \label{max1eq}
\sum_{u=1}^U \mathbb{E}[\ln p(\vec{m}(u) | \vec{w}(u), V)],
\end{equation}
where expectations are taken with respect to posterior i-vector distributions (similar to \eqref{kennyMaximization}). We propose maximizing the sum of log-likelihoods directly (\textbf{max. principle 2}) without computing expectations by setting $\vec{w}(u) = \vec{\mu}(u)$. That is, we maximize
\begin{align}
&\sum_{u=1}^U \ln p(\vec{m}(u) | \vec{w}(u), V) \label{max2eq} \\
&= \sum_{u=1}^U \Big( - \frac{h}{2} \ln(2\pi\sigma^2) \nonumber \\
&\phantom{\qquad} - \frac{1}{2\sigma^2} \big(\vec{m}(u) - V \vec{\mu}(u)\big)\T \big(\vec{m}(u) - V \vec{\mu}(u)\big) \Big) \nonumber \\
&= - \frac{h U}{2} \ln(2\pi\sigma^2) - \frac{1}{2\sigma^2} \sum_{u=1}^U \Big(
\vec{m}(u)\T \vec{m}(u) \nonumber \\
&\phantom{\qquad} - 2 \vec{m}(u)\T V \vec{\mu}(u) + \vec{\mu}(u)\T V\T V \vec{\mu}(u) \Big), \nonumber
\end{align}
where $h$ is the dimension of supervectors.

By taking derivatives with respect to each variable in the matrix $V$ and by setting them to zero, we obtain
\begin{empheq}[box={\textfbox[Update Formulas for $V$ and $\sigma^2$]}]{align*}
&V = \bigg( \sum_{u=1}^U 
\vec{m}(u) \vec{\mu}(u)\T \bigg) \bigg(\sum_{u=1}^U \mathbb{E}_{\mu\mu}(u) \bigg)^{-1}, \\
&\sigma^2 = \frac{1}{h U} \sum_{u=1}^U \Big(
\vec{m}(u)\T \vec{m}(u) - \Tr\big(\mathbb{E}_{\mu\mu}(u) V\T V \big)  \Big), \\
&\mathbb{E}_{\mu\mu}(u) = \vec{\mu}(u) \vec{\mu}(u)\T \quad \textrm{(\textbf{max principle 2})}. 
\end{empheq}

The traditional approach (\textbf{max principle 1}) of solving $V$ results in the exact same formulas but with 
\[
\mathbb{E}_{\mu\mu}(u) = \Sigma + \vec{\mu}(u) \vec{\mu}(u)\T \quad \textrm{\cite{tipping1999probabilistic}}. 
\]
Similarly to FEFA, training of $V$ is iterative. As initial values, we use random $V$ and $\sigma^2 = 1$.

\subsection{Factor Analysis}
\emph{Factor analysis} (FA) agrees with the model \eqref{ppcaModel} of PPCA, except that the noise term $\vec{\epsilon}$ has more freedom by letting \mbox{$\vec{\epsilon}~\sim~\mathcal{N}(\vec{0}, \Psi)$}, where $\Psi \in \R^{h \times h}$ is diagonal instead of isotropic~\cite{tipping1999probabilistic}. The training procedure is analogous to PPCA~\cite[pp. 585--586]{bishop2006machine}. First, the parameters of posterior distributions \eqref{posteriorDistribution} are computed as
\begin{empheq}[box={\textfbox[I-Vector Extraction]}]{align*}
\Sigma &= \Big(I + V\T \Psi^{-1} V\Big)^{-1}, \\
    \vec{\mu}(u) &= \Sigma V\T \Psi^{-1} \vec{m}(u).
\end{empheq}
Then, the model parameters are updated as follows:
\begin{empheq}[box={\textfbox[Update Formulas for $V$ and $\Psi$]}]{align*}
&V = \bigg( \sum_{u=1}^U 
\vec{m}(u) \vec{\mu}(u)\T \bigg) \bigg(\sum_{u=1}^U \mathbb{E}_{\mu\mu}(u) \bigg)^{-1}, \\
&\Psi = \frac{1}{U} \sum_{u=1}^U \Big( \vec{m}(u) \vec{m}(u)\T - V \mathbb{E}_{\mu\mu}(u) V\T \Big) \odot I,
\end{empheq}
where $\odot$ denotes the Hadamard (element-wise) product. The update formula for matrix $V$ is the same as with PPCA.

\subsection{Supervised Approaches}

\label{supervisedSection}

The methods presented so far capture the variability between individual utterances regardless of the speaker's identity. That is, their training is unsupervised in the sense that no speaker labels are needed. This makes it convenient to apply any of the above methods to different classification tasks, but leaves an open question whether ``better'' i-vectors could be extracted by utilizing speaker labels. Thus, we explore two methods that include identity information to the training process of the total variability matrix $V$ to discriminate speakers better. The explored methods, recently proposed \emph{probabilistic partial least squares} (PPLS)~\cite{chen2017speaker} and \emph{supervised PPCA} (SPPCA)~\cite{lei2010speaker}, can both be thought as extensions of the regular PPCA. The underlying models of PPLS and SPPCA are essentially the same, where the difference is in the data used to discriminate speakers: PPLS adds discrimination by using \emph{speaker labels} while SPPCA utilizes \emph{speaker-dependent sufficient statistics} within the FEFA framework. In the current work, however, we apply SPPCA in the supervector context in contrast to~\cite{lei2010speaker}, where the FEFA context is used.

In PPLS, supervector model is bundled together with a speaker label model. Speaker labels are encoded as \emph{one-hot vectors}, $\vec{y}(u) \in \R^s$, where $s$ is the number of speakers in the training set. For example, if the utterance $u$ originates from the second speaker of the set, then $\vec{y}(u) = (0, 1, 0, \ldots, 0)\T$. The speaker label model and the supervector model share the same i-vector $\vec{w}(u)$ as follows:
\begin{empheq}[left=\empheqlbrace]{align}
	\vec{m}(u) &= V\vec{w}(u) + \vec{\epsilon}, \phantom{\vec{\zeta}Q} \quad \textrm{(supervector model)} \label{pplsSuperVectorModel}\\
    \vec{y}(u) &= Q\vec{w}(u) + \vec{\zeta}, \phantom{\vec{\epsilon}V} \quad \textrm{(speaker label model)} \nonumber
\end{empheq}
where \eqref{pplsSuperVectorModel} is defined in the same way as before, $Q \in \R^{s \times d}$ is a mapping between the i-vector space and the one-hot vector space, and $\vec{\zeta} \sim \mathcal{N}(\vec{0},\rho^2 I)$.

As presented in~\cite{chen2017speaker} and~\cite{lei2010speaker}, the PPLS model leads to
\begin{empheq}[box={\textfbox[I-Vector Extraction]}]{align}
\label{pplsExtraction}
\begin{split}
\Sigma &= \Big(I + \frac{1}{\sigma^2}V\T V + \frac{1}{\rho^2}Q\T Q \Big)^{-1}, \\
    \vec{\mu}(u) &= \Sigma \Big( \frac{1}{\sigma^2}  V\T \vec{m}(u) + \frac{1}{\rho^2}  Q\T \vec{y}(u)\Big)
    \end{split}
\end{empheq}
and
\begin{empheq}[box={\textfbox[Update Formulas for $V$, $Q$, $\sigma^2$, and $\rho^2$ ]}]{align}
&V = \bigg( \sum_{u=1}^U 
\vec{m}(u) \vec{\mu}(u)\T \bigg) \bigg(\sum_{u=1}^U \mathbb{E}_{\mu\mu}(u) \bigg)^{-1}, \nonumber \\
&Q = \bigg( \sum_{u=1}^U 
\vec{y}(u) \vec{\mu}(u)\T \bigg) \bigg(\sum_{u=1}^U \mathbb{E}_{\mu\mu}(u) \bigg)^{-1}, \nonumber \\
&\sigma^2 = \frac{1}{h U} \sum_{u=1}^U \Big(
\vec{m}(u)\T \vec{m}(u) - \Tr\big(\mathbb{E}_{\mu\mu}(u) V\T V \big)  \Big), \nonumber \\
&\rho^2 = \frac{1}{s U} \sum_{u=1}^U \Big(
\vec{y}(u)\T \vec{y}(u) - \Tr\big(\mathbb{E}_{\mu\mu}(u) Q\T Q \big)  \Big), \label{rhoUpdate}
\end{empheq}
where, as before, we assume that all the supervectors and one-hot vectors are centered using the mean vectors calculated from the training data.

To extract i-vectors as in \eqref{pplsExtraction}, we are required to have speaker information of the utterance stored in $\vec{y}(u)$. In the testing phase, however, we have no information about the speaker. To this end, we might simply extract the test i-vector using extraction formulas \eqref{ppcaExtraction} for PPCA~\cite{lei2010speaker}. The result is not the same as with PPCA, since the training of $V$ is influenced by the supervised approach. Another solution to deal with the lacking speaker information is to predict speaker labels as a mean of posterior distribution $p(\vec{y}(u) | \vec{m}(u))$; see the details in~\cite{chen2017speaker}. We found experimentally that both approaches extract exactly the same i-vectors.

The described formulations apply also for SPPCA with a difference that instead of using one-hot encoded speaker labels as vectors $\vec{y}(s)$, we use speaker dependent supervectors. A speaker dependent supervector is formed by using the acoustic features from all of the speaker's training utterances. Note that with SPPCA, $h$ should be used instead of $s$ in \eqref{rhoUpdate} and that $Q$'s dimensionality is the same as $V$'s.

In~\cite{lei2010speaker}, a \emph{weighted} SPPCA is proposed. In weighted SPPCA,  \eqref{pplsExtraction} becomes
\begin{empheq}[box={\textfbox[I-Vector Extraction]}]{align*}
\begin{split}
\Sigma &= \Big(I + \frac{1}{\sigma^2}V\T V + \frac{\beta}{\rho^2}Q\T Q \Big)^{-1}, \\
    \vec{\mu}(u) &= \Sigma \Big( \frac{1}{\sigma^2}  V\T \vec{m}(u) + \frac{\beta}{\rho^2}  Q\T \vec{y}(u)\Big),
    \end{split}
\end{empheq}
where $\beta$ is a weight parameter. The weight parameter can be used to adjust the amount of supervision added on top of the conventional PPCA model. With $\beta = 0$, the model equals PPCA and when $\beta = 1$, we have the ordinary SPPCA.


\section{Experimental Setup}

\subsection{VoxCeleb Speaker Recognition Corpus}
We performed the speaker verification experiments on the recently published \emph{VoxCeleb} dataset~\cite{nagrani2017voxceleb}. VoxCeleb contains over 150,000 real-world utterances from 1251 celebrities, of which 561 are females and 690 are males. A key difference to the widely used NIST corpora is that, on average, VoxCeleb has more than 100 utterances per speaker, typically obtained from multiple sessions with highly variable environments and recording conditions providing a large intra-speaker variability. The average utterance length is about 8 seconds. Although most of the utterances are short, the utterance length varies considerably, the longest ones being longer than one minute. Utterances have a sampling rate of 16 kHz.

The dataset was collected using fully automated pipeline that extracts and annotates utterances from YouTube videos. To ensure correct speaker annotation, the pipeline contains automatic face verification verifying that mouth movement in the video corresponds to the audio track and that the speaker's identity is the correct one. The utterances are mostly extracted from interview situations ranging from quiet studios to public speeches in front of large audiences. Differing environments and speaking styles are not the only source of variability, since differences in recording devices and audio processing practices are present in YouTube videos. As the acoustic and technical conditions of the utterances vary considerably, the dataset turns out to be challenging for an automatic speaker verification task as we will see in the experiments.

We used the same standard trial list as in the baseline system of~\cite{nagrani2017voxceleb}. It contains 40 speakers, whose name starts with the letter `E'.  The list has 37720 trials, half of them (18860) being same speaker trials, which differs substantially from the typical NIST setups with about 10 to 1 ratio between non-target and target trials.

The rest of the speakers were used for developing our speaker verification systems, \ie to train the UBM, TVM and classifier back-end. To speed up experimentation, we utilized only one-fourth of the available training data as we did not see large decrease in system performance by decreasing the amount of training data. Our training set contains total of 37160 utterances from 1211 speakers.

We report recognition accuracy using \emph{equal error rates} (EERs) that are the rates where \emph{false alarm} and \emph{miss} rates are equal. With the current trial list setup, 95\% confidence intervals around EERs that are below 8\% (the case with most experiments) are at widest $\pm 0.27$\% absolute. Confidence intervals are computed using \emph{z-test} based methodology presented in~\cite{bengio2004statistical}.

\subsection{NIST Data}
Even if our primary interest is in the VoxCeleb data, for the sake of reference, we also study different i-vector systems using common condition 5 of NIST 2010 Speaker Recognition Evaluation (SRE10)\footnote{https://www.nist.gov/itl/iad/mig/speaker-recognition-evaluation-2010}. Trial segments in condition 5 contain conversational 8 kHz telephone speech spoken with normal vocal effort. The trial list consists of 30373 trials, of which 708 are same speaker trials.

Speaker verification systems were trained using 43308 utterances obtained from SRE04, SRE05, SRE06, Switchboard, and Fisher corpora. 

The performances are reported as EERs and \emph{minimum values} of \emph{detection cost function} (minDCF) used in SRE10. The SRE10 detection cost function is given as
\[
	\textrm{DCF} = 0.001 \, P_{\textrm{miss}} + 0.999 \, P_{\textrm{fa}},
\]
where $P_{\textrm{miss}}$ and $P_{\textrm{fa}}$ are probabilities of miss and false alarm, respectively~\cite{martin2010nist}.


\subsection{Description of Speaker Verification System}
We extracted 38 dimensional acoustic feature vectors containing 19 \emph{Mel-frequency cepstral coefficients} (MFCCs) and their delta coefficients. After discarding features of non-speech segments, we subjected features to utterance-level mean and variance normalization.

The \emph{universal background model} (UBM), 1024 component \emph{Gaussian mixture model} (GMM) with diagonal covariance matrices, was trained using the development data. UBM was used to extract sufficient statistics, which were used in FEFA or in supervector extraction. Supervectors were extracted by first creating utterance specific GMMs using \emph{maximum a posteriori} (MAP) adaptation~\cite{reynolds2000speaker} and then by concatenating mean vectors of adapted GMMs into supervectors.

We extracted 400 dimensional i-vectors, which were then centered, length-normalized, and whitened. Finally, we used simplified \emph{probabilistic linear discriminant analysis} (PLDA)~\cite{garcia2011analysis} to perform supervised dimensionality reduction of i-vectors into 200 dimensions and to score verification trials.

\section{Speaker Verification Experiments}

The results presented in Sections \ref{relevanceFactorSection} to \ref{dimensionalitySection} are given for the VoxCeleb ASV protocol. Section \ref{nistResultsSection} presents results for SRE10.

\subsection{Relevance Factor in MAP Adaptation}
\label{relevanceFactorSection}

\begin{figure}[h]
	\centering
	\includegraphics[width=0.9\linewidth]{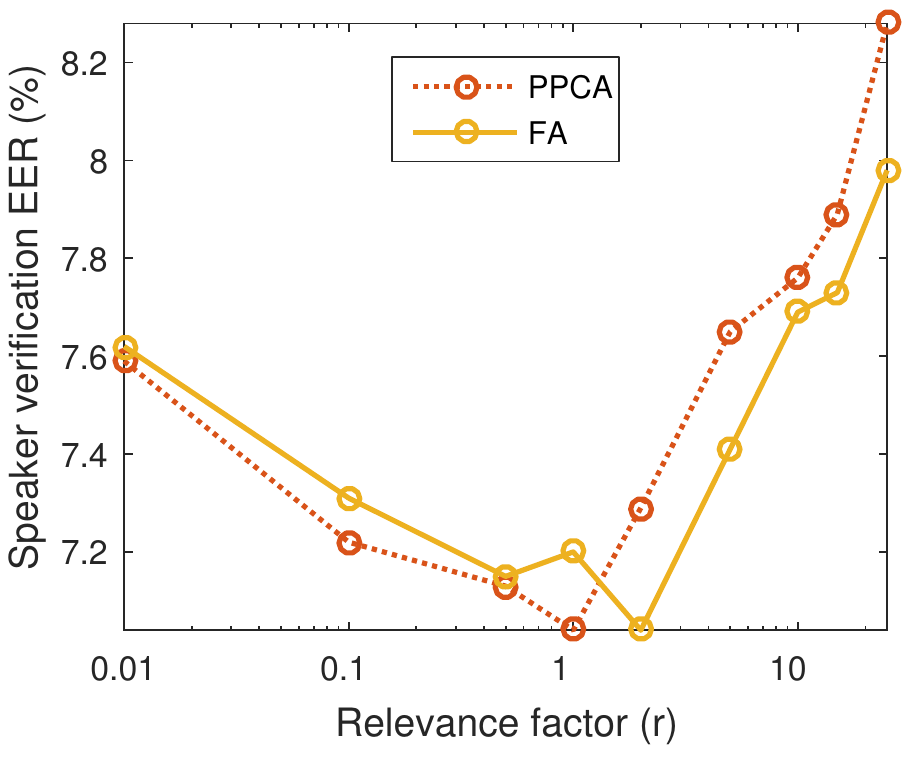}
    \vspace{-3mm}
	\caption{{\it The effect of relevance factor used in MAP adaptation on speaker verification performance. For VoxCeleb data, relevance factor close to 1 leads to the best results.}}
	\label{relevanceFactorFigure}
\end{figure}

We studied how the choice of relevance factor affects speaker verification performance with PPCA and FA methods. The results for VoxCeleb protocol are presented in Figure \ref{relevanceFactorFigure}. Based on the results, we fix the relevance factor to $r = 1$ for the remaining experiments with this data. The choice of relevance factor is data-dependent, and therefore, the same value might not work well with other datasets.

\subsection{Number of Training Iterations}

\begin{figure}[h]
	\centering
	\includegraphics[width=0.9\linewidth]{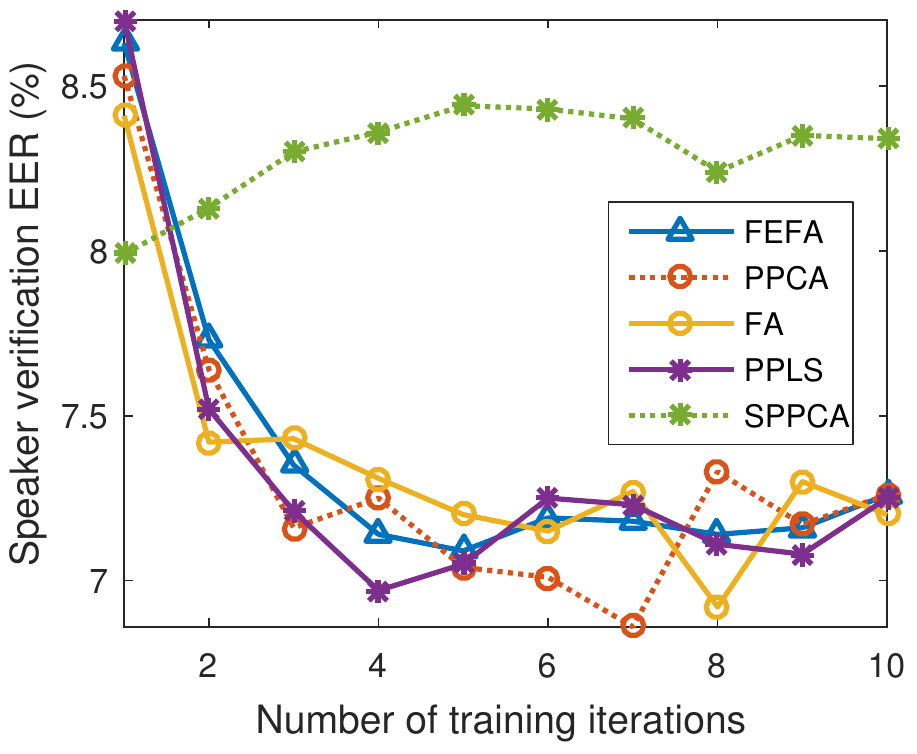}
    \vspace{-3mm}
	\caption{{\it Speaker verification performances with different numbers of training iterations in total variability model training. About 5 iterations are enough to obtain satisfying performance. The trend is similar for all the methods except for SPPCA, for which the iterative training does not improve the results.}}
	\label{iterationsAndMethodsFigure}
\end{figure}

To find out the sufficient number of iterations in TVM training, we evaluated verification performances with varying number of iterations. The results of the experiment, presented in Figure \ref{iterationsAndMethodsFigure}, reveal that 5 iterations are enough to obtain near to optimal performance. Hence, we fix the number of iterations to 5 for the remaining experiments. 

All the methods, except SPPCA, behave similarly. With SPPCA, the training does not proceed in a desirable way during the first 5 iterations.

\subsection{Maximization Principles in System Training}
In Section \ref{ppcaSection}, we presented two different maximization principles that can be used with all the discussed iterative TVM training methods. The comparison of the maximization principles in terms of resulting speaker verification performances is presented in Table \ref{maxPrinciplesTable}. There are no clear differences between the principles.

\begin{table}[h]
 \caption{\label{maxPrinciplesTable} {\it Speaker verification equal error rates (\%) for different methods and maximization principles used in TVM training. With conventional PCA approach we obtained EER of 7.39\%.}}
 \vspace{2mm}
 \centerline{
 \begin{tabular}{l c c}
 \hline
& max1 [Eq. \eqref{max1eq}] & max2 [Eq. \eqref{max2eq}] \\
\hline
FEFA & 7.09 & 7.11 \\
PPCA & \textbf{7.04} & 7.18 \\
FA & 7.20 & 7.26 \\
PPLS & 7.05 & 7.42 \\
SPPCA & 8.44 & 8.26 \\
 \hline
 \end{tabular}}
 \end{table}

\subsection{Training Times}

\begin{figure}[h]
	\centering
	\includegraphics[width=0.9\linewidth]{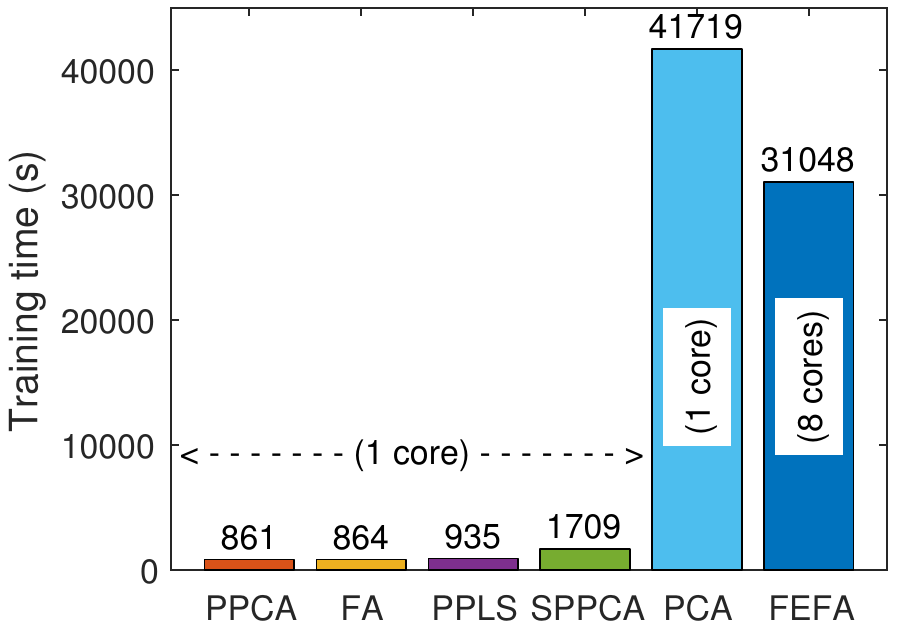}
    \vspace{-3mm}
	\caption{{\it Training times of TVMs with 5 iterations (PCA is non-iterative). Iterative supervector compression methods are the fastest to train, while FEFA requires the most amount of computation. FEFA was trained with 8 CPU cores to reduce the training time.}}
	\label{trainingTimesFigure}
\end{figure}

Figure \ref{trainingTimesFigure} shows the elapsed times for 5 TVM training iterations with different methods. The measured times do not include the times needed to compute sufficient statistics or supervectors or to load them into memory using I/O operations. Measurements were conducted by running MATLAB implementations of all the methods in a 16-core 2.60 GHz machine with an ample amount of RAM ($>$300GB). To obtain reasonable training time with FEFA, we trained the system with 8 CPU cores, whereas other methods were trained using a single core.

Before the TVM training phase, different methods have only small differences in terms of computational requirements. Even though FEFA differs from other methods in that it uses sufficient statistics as inputs, the difference is minuscule, as most of the time in the extraction of MAP adapted supervectors is spent in the computation of sufficient statistics.

From the perspective of system optimization, note that FEFA does not require optimization of the relevance factor. But, the extra cost of relevance factor optimization in PPCA-PLDA system does not outweigh the training time of FEFA, as MAP adaptation using precomputed sufficient statistics and training of PPCA and PLDA are much less expensive operations than FEFA training.

\subsection{Weight Parameter in Supervised Approaches}

\begin{figure}[h]
	\centering
	\includegraphics[width=0.9\linewidth]{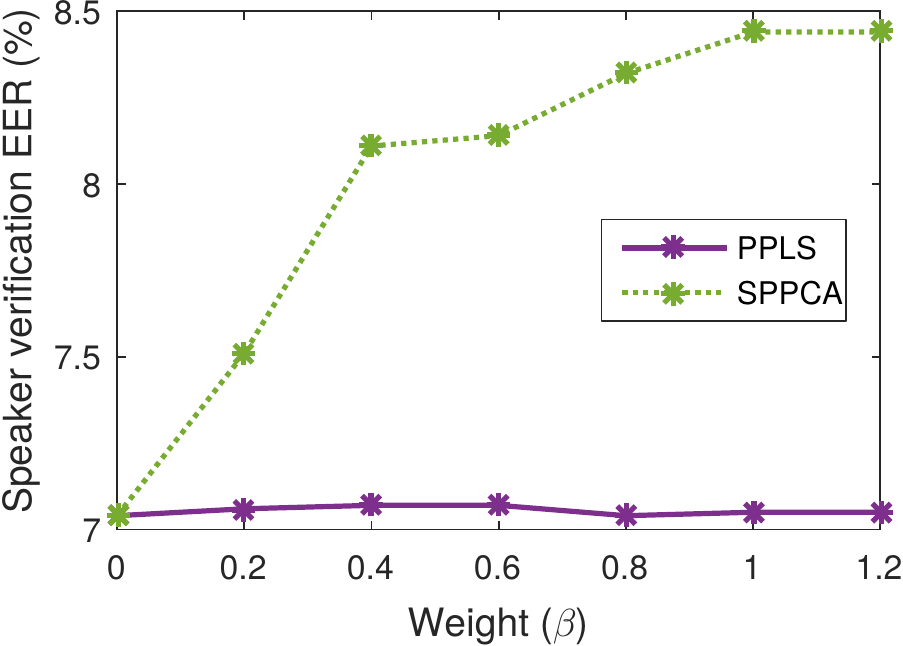}
    \vspace{-3mm}
	\caption{{\it Effect of the weight parameter ($\beta$) value in the supervised models. When $\beta = 0$, both models equal to the conventional PPCA. Adding supervision by increasing $\beta$ does not improve the speaker verification performance.}}
	\label{weightedModelsFigure}
\end{figure}

Next, we apply weighting to the supervised methods, PPLS and SPPCA, as discussed in Section \ref{supervisedSection}. The results obtained with different weight parameter values are presented in Figure~\ref{weightedModelsFigure}. We find that weighting does not affect PPLS and that the weighted SPPCA model functions better when it approaches PPCA \mbox{($\beta \to 0$)}. This suggests that the studied supervised methods do not provide any noticeable benefits over the unsupervised i-vector extractors.

\subsection{Dimensionality of I-Vectors}
\label{dimensionalitySection}

\begin{figure}[h]
	\centering
	\includegraphics[width=0.9\linewidth]{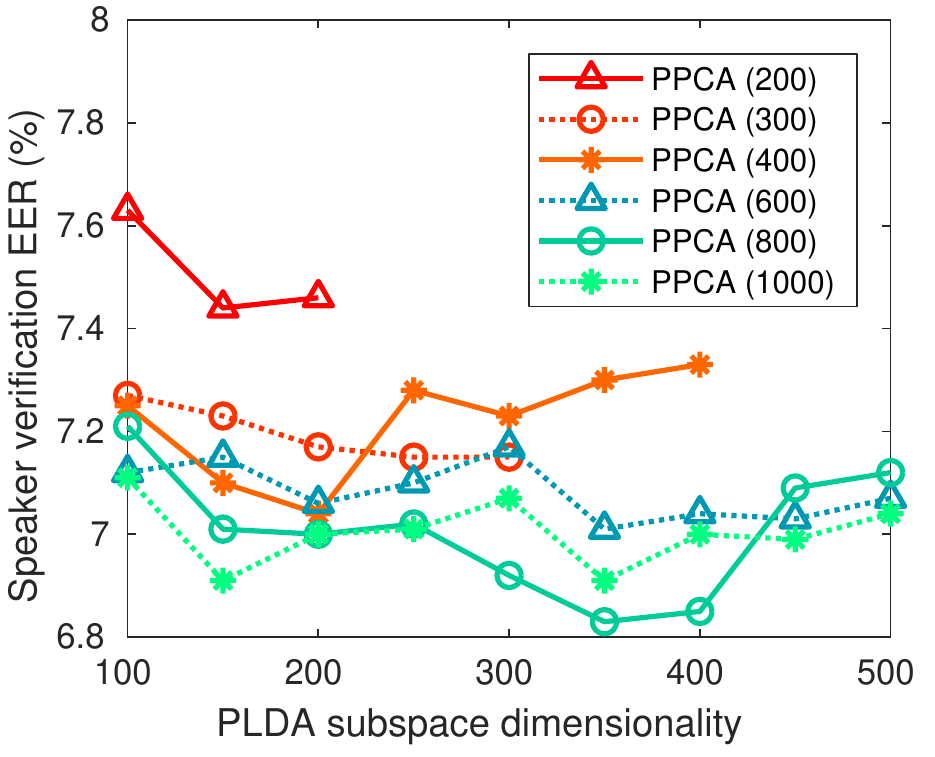}
    \vspace{-3mm}
	\caption{{\it Optimization of i-vector and PLDA subspace dimensions for PPCA method. Different lines represent different i-vector dimensionalities. The lowest error rate is obtained with 800-dimensional i-vectors and 350-dimensional PLDA space.}}
	\label{dimensionalityAndPldaFigure}
\end{figure}

To improve the speaker verification performance, we jointly optimized dimensions of i-vectors and their PLDA-compressed versions. We varied the i-vector dimensionality between 200 and 1000 and the PLDA subspace dimensionality between 100 and 500. The results for all combinations using PPCA method are presented in Figure \ref{dimensionalityAndPldaFigure}. The results indicate that, our initial parameters, 400 and 200 for i-vectors and PLDA, respectively, give a relatively good performance. We also see that a slight increase in performance might be obtained by using i-vector dimensions between 600 and 1000 with 350 to 400 dimensional PLDA subspaces.

\subsection{Evaluation With NIST SRE10 Data}

\label{nistResultsSection}

To increase our confidence of the generality of the results, we tested different i-vector systems on common condition 5 of SRE10. We ran the second protocol using mostly the same system configuration as with VoxCeleb corpus. We changed the filterbank spacing in MFCCs to match the 8 kHz sampling rate and we also increased the relevance factor to 6. Otherwise, the system was not optimized for the new data.

\begin{figure}[t]
	\centering
	\includegraphics[width=0.9\linewidth]{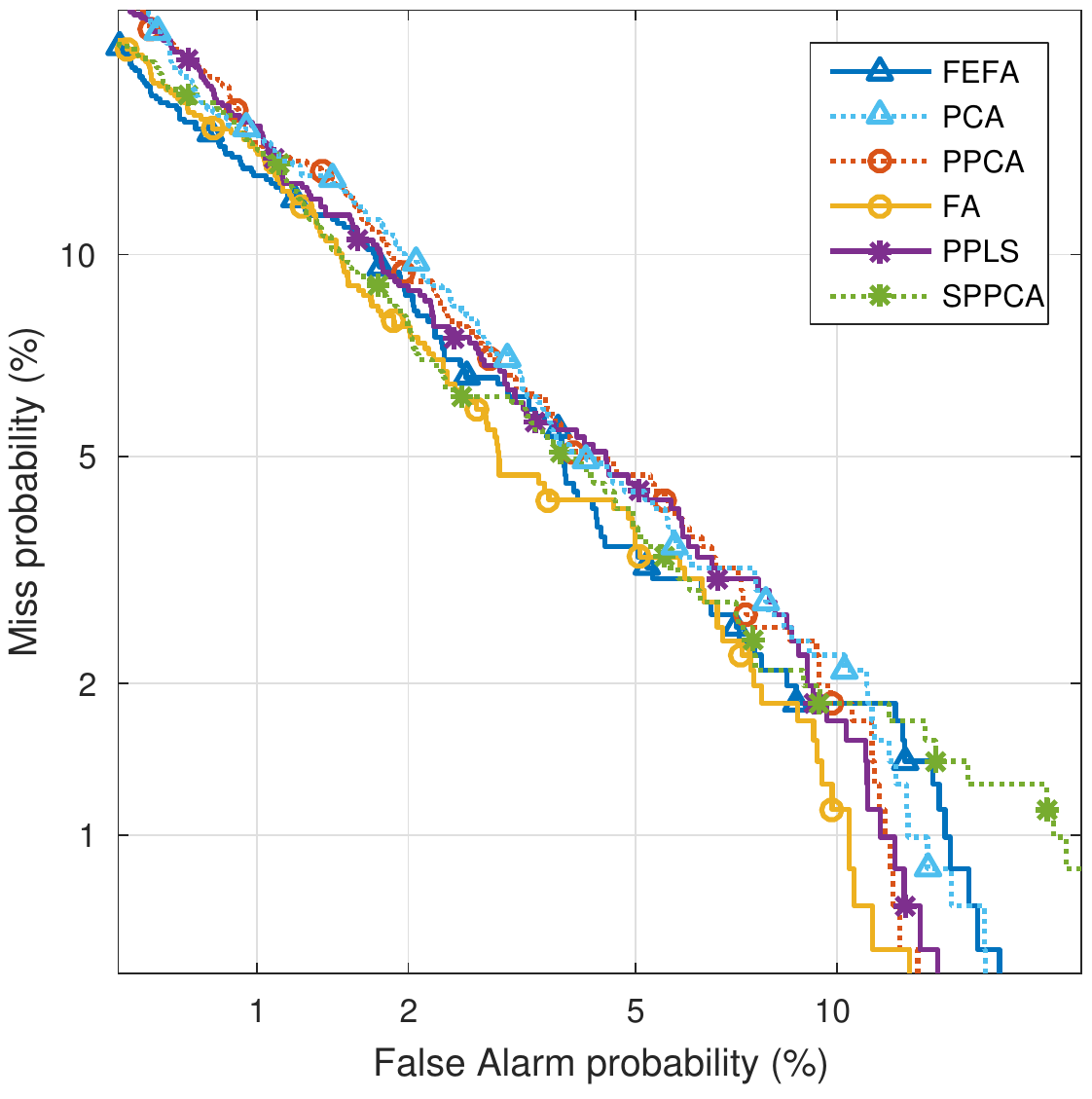}
    \vspace{-3mm}
	\caption{{\it Detection error trade-off curves for different methods using max. principle 1 on common condition 5 of SRE10.}}
	\label{detPlotFigure}
\end{figure}

The results for SRE10, shown in Table \ref{sre10table}, indicate that there are no clear differences between the two maximization principles. Further, we observe that the results for EER and minDCF are somewhat different as FEFA performs the best in terms of EER, while FA obtains the lowest minDCF. To gain better understanding on the performances of the systems, in Figure \ref{detPlotFigure}, we present \emph{detection error trade-off} (DET) curves for all the methods using the maximization principle 1. The DET curves reveal that there are no clear differences between FEFA and FA.

\section{Discussion and Conclusions}

The development and optimization of i-vector systems tends to be time consuming. In particular, any change in the acoustic front-end or the UBM configuration requires retraining the TVM. If TVM training is slow, the parameter optimization can become unfeasible, possibly leading to suboptimal system configuration. In this work, we studied fast GMM supervector compression methods to streamline ASV system development. By focusing on compression of MAP adapted supervectors, we managed to cut the system training time down to a fraction of traditional approach (FEFA).
Our results indicate that the alternative approaches work as well as the standard FEFA in terms of recognition accuracy. 
The less-optimistic performance reported in~\cite{madikeri2014fast} and~\cite{madikeri2012hybrid} (for the PPCA system) could be due to absence of MAP adaptation: we found that increasing the relevance factor from 0 (no MAP adaptation) towards some higher (optimized) value results in considerably higher verification accuracy. 

We did not find benefit with either of the studied supervised models, PPLS or SPPCA. We were not, therefore, able to reproduce positive findings claimed in~\cite{chen2017speaker} for PPLS. This might be due to differing datasets or feature configurations. On a positive side, we found that PPLS attains similar training speeds to PPCA and FA.

The proposed modification to the maximization principle in TVM training did not affect verification results negatively. This modification makes the theory and the system implementation slightly simpler as it avoids computing expectations over i-vector posterior distributions. 

We recognize that the findings of the current study can not be generalized to all existing system configurations without further studies. In this study, we only experimented with a specific set of acoustic features together with a specific UBM and back-end configurations (PLDA).

In summary, from the various compared variants, we recommend to use PPCA and FA to compress supervectors. Both are easy to implement on top of the GMM framework and lead to considerably faster TVM training times. For optimal verification accuracy, supervectors should be created using the MAP adaptation with an optimized relevance factor. We have made our MATLAB implementations of PPCA, FA, PPLS, and SPPCA available at \url{http://cs.uef.fi/~vvestman/research_codes/supervector_compression.zip}.

\begin{table}[t]
 \caption{\label{sre10table} {\it Speaker verification performances for different methods and maximization principles (max1, max2) on common condition 5 of SRE10. With conventional PCA we obtained EER of 4.69\% and minDCF of 6.55\%.}}
 \vspace{2mm}
 \centerline{
 \begin{tabular}{| l | c c | c c |}
  \hline
& \multicolumn{2}{c|}{EER (\%)} & \multicolumn{2}{c|}{minDCF (\%)} \\
 \cline{2-5}
& max1 & max2 & max1 & max2 \\
\hline
FEFA & 4.24 & \textbf{3.86} & 6.24 & 6.85 \\
PPCA & 4.66 & 4.66 & 6.55 & 6.72 \\
FA & 4.24 & 4.24 & 6.00 & \textbf{5.50} \\
PPLS & 4.66 & 4.79 & 6.73 & 6.53 \\
SPPCA & 4.46 & 4.38 & 6.46 & 6.26 \\
 \hline
 \end{tabular}}
 \end{table}

\bibliographystyle{IEEEbib}
%
\bibliography{mybib}

\appendix 

\section{Proof: I-Vector Posterior Distribution for PPCA Model}
The following proof is similar in principle to the proof of Proposition 1 in~\cite{kenny2005eigenvoice}.

It is enough to show that $p(\vec{w}(u)|\vec{m}(u)) \propto \mathcal{N}(\vec{\mu}(u), \Sigma)$, since $p(\vec{w}(u)|\vec{m}(u))$ is a probability density function and will hence be correctly scaled. For brevity, we drop $u$ from the notation of the following chain of relations that proves the claim (\eg $\vec{\mu}$ will refer to $\vec{\mu}(u)$):
\label{posteriorProof}
\begin{align*}
&\mathcal{N}(\vec{\mu}, \Sigma) \\
&\propto \exp \Big(-\frac {1}{2} (\vec{w} - \vec{\mu})\T \Sigma^{-1} (\vec{w} - \vec{\mu}) \Big) \\
&= \exp \Big(-\frac {1}{2} \Big(\vec{w} - \frac{1}{\sigma^2} \Sigma V\T \vec{m}\Big)\T \Big(I + \frac{1}{\sigma^2}V\T V\Big) \\
&\phantom{\qquad} \Big(\vec{w} - \frac{1}{\sigma^2} \Sigma V\T \vec{m}\Big) \Big) \\
&= \exp \Big(-\frac {1}{2} \Big[ \Big(\vec{w}\T 
+ \frac{1}{\sigma^2} \vec{w}\T V\T V 
- \frac{1}{\sigma^2} (\Sigma V\T \vec{m})\T \Sigma^{-1} \Big) \\
&\phantom{\qquad} \Big(\vec{w} - \frac{1}{\sigma^2} \Sigma V\T \vec{m}\Big) \Big] \Big) \\
&= \exp \Big(-\frac {1}{2} \Big[ \vec{w}\T \vec{w}
+ \frac{1}{\sigma^2} \vec{w}\T V\T V \vec{w} \\
&\phantom{\qquad} - \frac{1}{\sigma^2} \Big(\Sigma V\T \vec{m} \Big)\T \Sigma^{-1} \vec{w} \\
&\phantom{\qquad} - \Big(\vec{w}\T + \frac{1}{\sigma^2} \vec{w}\T V\T V \Big) \frac{1}{\sigma^2} \Sigma V\T \vec{m} \Big] + \textrm{const}(\vec{m}) \Big) \\
&\propto \mathcal{N}(\vec{0},I) \exp \Big(-\frac {1}{2\sigma^2} \Big[
 (V \vec{w})\T V \vec{w}
- \vec{m}\T V \Sigma\T \Sigma^{-1} \vec{w} \\
&\phantom{\qquad} - \vec{w}\T \Sigma^{-1} \Sigma V\T \vec{m} \Big] \Big) \\
&\propto \mathcal{N}(\vec{0},I) \exp \Big(-\frac {1}{2\sigma^2} \Big[
 (V \vec{w})\T V \vec{w}
- \vec{m}\T V \vec{w} \\
&\phantom{\qquad} - (V \vec{w})\T \vec{m} + \vec{m}\T \vec{m} \Big] \Big) \\
&= \mathcal{N}(\vec{0},I) \exp \Big(-\frac {1}{2} (\vec{m} - V \vec{w})\T \frac{1}{\sigma^2} (\vec{m} - V \vec{w})\Big) \\
&\propto \frac{p(\vec{w}) p(\vec{m}|\vec{w})}{p(\vec{m})} \\
&= p(\vec{w}|\vec{m}).
\end{align*}
Note that in the above chain of proportional relations, we can drop ($\exp(\textrm{const}(\vec{m}))$) and add ($\exp(\vec{m}\T \vec{m})$; $p(\vec{m})$) multipliers that only depend on $\vec{m}$ without breaking the chain.

\end{document}